\documentclass[dvips]{article}
\usepackage{supertabular,lscape,epsfig}
\usepackage{amssymb}
\usepackage{amsmath}
\DeclareSymbolFont{ppa}{OT1}{ppl}{m}{it}
\DeclareMathSymbol{\vv}{\mathalpha}{ppa}{'166}

\begin{document}

\newcommand{\dd}{\,{\rm d}}
\newcommand{\ie}{{\it i.e.},\,}
\newcommand{\etal}{{\it et al.\ }}
\newcommand{\eg}{{\it e.g.},\,}
\newcommand{\cf}{{\it cf.\ }}
\newcommand{\vs}{{\it vs.\ }}
\newcommand{\zdot}{\makebox[0pt][l]{.}}
\newcommand{\up}[1]{\ifmmode^{\rm #1}\else$^{\rm #1}$\fi}
\newcommand{\dn}[1]{\ifmmode_{\rm #1}\else$_{\rm #1}$\fi}
\newcommand{\upd}{\up{d}}
\newcommand{\uph}{\up{h}}
\newcommand{\upm}{\up{m}}  
\newcommand{\ups}{\up{s}}
\newcommand{\arcd}{\ifmmode^{\circ}\else$^{\circ}$\fi}
\newcommand{\arcm}{\ifmmode{'}\else$'$\fi}
\newcommand{\arcs}{\ifmmode{''}\else$''$\fi}
\newcommand{\MS}{{\rm M}\ifmmode_{\odot}\else$_{\odot}$\fi}
\newcommand{\RS}{{\rm R}\ifmmode_{\odot}\else$_{\odot}$\fi}
\newcommand{\LS}{{\rm L}\ifmmode_{\odot}\else$_{\odot}$\fi}

\newcommand{\Abstract}[2]{{\footnotesize\begin{center}ABSTRACT\end{center}
\vspace{1mm}\par#1\par   
\noindent
{~}{\it #2}}}

\newcommand{\TabCap}[2]{\begin{center}\parbox[t]{#1}{\begin{center}
  \small {\spaceskip 2pt plus 1pt minus 1pt T a b l e}
  \refstepcounter{table}\thetable \\[2mm]
  \footnotesize #2 \end{center}}\end{center}}

\newcommand{\TableSep}[2]{\begin{table}[p]\vspace{#1}
\TabCap{#2}\end{table}}

\newcommand{\FigCap}[1]{\footnotesize\par\noindent Fig.\  %
  \refstepcounter{figure}\thefigure. #1\par}

\newcommand{\TableFont}{\footnotesize}
\newcommand{\TableFontIt}{\ttit}
\newcommand{\SetTableFont}[1]{\renewcommand{\TableFont}{#1}}

\newcommand{\MakeTable}[4]{\begin{table}[htb]\TabCap{#2}{#3}
  \begin{center} \TableFont \begin{tabular}{#1} #4
  \end{tabular}\end{center}\end{table}}

\newcommand{\MakeTableSep}[4]{\begin{table}[p]\TabCap{#2}{#3}
  \begin{center} \TableFont \begin{tabular}{#1} #4
  \end{tabular}\end{center}\end{table}}

\newenvironment{references}%
{
\footnotesize \frenchspacing
\renewcommand{\thesection}{}
\renewcommand{\in}{{\rm in }}
\renewcommand{\AA}{Astron.\ Astrophys.}
\newcommand{\AAS}{Astron.~Astrophys.~Suppl.~Ser.}
\newcommand{\ApJ}{Astrophys.\ J.}
\newcommand{\ApJS}{Astrophys.\ J.~Suppl.~Ser.}
\newcommand{\ApJL}{Astrophys.\ J.~Letters}
\newcommand{\AJ}{Astron.\ J.}
\newcommand{\IBVS}{IBVS}
\newcommand{\PASP}{P.A.S.P.}
\newcommand{\Acta}{Acta Astron.}
\newcommand{\MNRAS}{MNRAS}
\renewcommand{\and}{{\rm and }}
\section{{\rm REFERENCES}}
\sloppy \hyphenpenalty10000
\begin{list}{}{\leftmargin1cm\listparindent-1cm
\itemindent\listparindent\parsep0pt\itemsep0pt}}%
{\end{list}\vspace{2mm}}
 
\def\TYLDA{~}
\newlength{\DW}
\settowidth{\DW}{0}
\newcommand{\dw}{\hspace{\DW}}

\newcommand{\refitem}[5]{\item[]{#1} #2%
\def\REFARG{#3}\ifx\REFARG\TYLDA\else, {\it#3}\fi
\def\REFARG{#4}\ifx\REFARG\TYLDA\else, {\bf#4}\fi
\def\REFARG{#5}\ifx\REFARG\TYLDA\else, {#5}\fi.}

\newcommand{\Section}[1]{\section{#1}}
\newcommand{\Subsection}[1]{\subsection{#1}}
\newcommand{\Acknow}[1]{\par\vspace{5mm}{\bf Acknowledgments.} #1}
\pagestyle{myheadings}

\newfont{\bb}{ptmbi8t at 12pt}
\newcommand{\xrule}{\rule{0pt}{2.5ex}}  
\newcommand{\xxrule}{\rule[-1.8ex]{0pt}{4.5ex}}  
\def\thefootnote{\fnsymbol{footnote}}
\begin{center}

{\Large\bf
CURiuos Variables Experiment (CURVE):
Variable Stars in the metal-poor Globular Cluster M56}
\vskip1cm
{\bf
P.~~P~i~e~t~r~u~k~o~w~i~c~z$^{1,2}$,~~ A.~~~O~l~e~c~h$^1$,
~~P.~~K~\c{e}~d~z~i~e~r~s~k~i$^3$,\\
~~K.~~Z~{\l}~o~c~z~e~w~s~k~i$^1$,
~~M.~~W~i~\'s~n~i~e~w~s~k~i$^1$,~~K.~~M~u~l~a~r~c~z~y~k$^3$\\}
\vskip3mm
{
  $^1$Nicolaus Copernicus Astronomical Center,
     ul. Bartycka 18, 00-716 Warsaw, Poland\\
     e-mail: (pietruk,olech,kzlocz,mwisniew)@camk.edu.pl\\
  $^2$Departamento de Astronom\'ia y Astrof\'isica, Pontificia
     Universidad Cat\'olica de Chile, Av. Vicu\~na MacKenna 4860,
     Casilla 306, Santiago 22, Chile \\
  $^3$Warsaw University Observatory,
     Al. Ujazdowskie 4, 00-478 Warsaw, Poland\\
     e-mail: pkedzier,kmularcz@astrouw.edu.pl\\
}
\end{center}

\Abstract{
We have surveyed a $6\zdot\arcm5 \times 6\zdot\arcm5$ field centered
on the globular cluster M56 (NGC 6779) in search for variable stars.
We have detected seven variables, among which two objects are new
identifications. One of the new variables is an RR Lyrae
star, the third such star in M56. Comparison of the new observations
and old photometric data for an RV Tauri variable V6 indicates
a likely period change in the star. Its slow and negative rate
of $-0.005\pm0.003$~d/yr would disagree with post-AGB
evolution, however this could be a result of blue-loop evolution
and/or random fluctuations of the period.
}
{Hertzsprung-Russell (HR) and C-M diagrams --
Stars: variables : BL Her, RV Tau, RR Lyr --
open clusters and associations: individual: M56 (NGC 6779)}

\Section{Introduction}

CURious Variables Experiment (CURVE) is a long-term project focused
on observations of open clusters, globular clusters and cataclysmic
variable stars in the northern hemisphere (Olech \etal 2003,
Olech \etal 2007, Rutkowski \etal 2007). In stellar clusters we
principally search for variable objects. However, our data also
allows us to estimate basic parameters of observed clusters, such as
distances and ages (Pietrukowicz \etal 2006).

The globular cluster M56 (NGC 6779) is located in a rather dense
galactic field at $(l,b)=(62\zdot\arcd66,+8\zdot\arcd34)$.
The most recent deep $BVRI$ photometry of the cluster
was obtained by Hatzidimitriou \etal (2004) using 1.3-m telescope
at Skinakas Observatory, in Crete. They estimate the distance
modulus and the reddening for M56 of $(m-M)_V=15.62\pm0.26$
and $E(B-V)=0.32\pm0.02$, respectively. The authors also demonstrate
that M56 is one of the most metal-poor ([Fe/H]$_{CG}=-2.00\pm0.21$
on the scale proposed by Carretta and Gratton 1997)
and one of the oldest globular clusters in the Galactic halo
(13 Gyrs, using the age-index calibration of Salaris and Weiss 2002).

Despite the very early discovery of the first variable star in the
globular cluster M56 (the object classified now as V3, Davis 1917) and
excellent position for northern hemisphere observers of the cluster,
identification of its variables has proceeded very slowly.
Clement \etal (2001) lists only 12 variable stars in M56,
but five of them are very likely field objects.
In this contribution we present results of the search
for variable stars in M56 based on new data and
with the use of image subtraction method, which works much
better in crowded fields than classical photometry.

\Section{Observations and Data Reductions}

Observations of the globular cluster M56 were made during
48 nights between July 5, 2002 and May 25, 2004 at the Ostrowik
station of the Warsaw University Observatory. The data were collected
using the 60-cm Cassegrain telescope equipped with a Tektronics
TK512CB back-illuminated CCD camera. The scale of the camera was
$0\zdot\arcs76$/pixel providing a $6\zdot\arcm5 \times 6\zdot\arcm5$
field of view. The full description of the telescope and camera
was given by Udalski and Pych (1992).

We monitored the cluster in ``white light'', which roughly corresponds
to the Cousins $R$ band (Udalski and Pych 1992). The exposure times
were from 120 to 240 seconds. We analyzed 543 images
with seeing better than $4\zdot\arcs56$ ($<6.0$ pixels) and average
background level lower than 3500. The best measured seeing reached
$2\zdot\arcs90$. Table~1 lists the nights during which the data
were obtained.

\begin{table}
\centering
\caption{\small Dates of observations of M56}
\medskip
{\small
\begin{tabular}{lll}
\hline
Year & Month     & Nights \\
\hline
2002 & July      & 5/6, 6/7, 8/9, 9/10, 13/14 \\
2002 & August    & 1/2, 10/11, 11/12, 13/14 \\
2003 & May       & 24/25, 25/26 \\
2003 & June      & 2/3, 24/25, 26/27, 30/31 \\
2003 & August    & 17/18, 18/19, 20/21, 22/23, 24/25, \\
     &           & 25/26, 27/28, 28/29, 29/30, 30/31, 31/1 \\
2003 & September & 1/2, 2/3, 3/4, 5/6, 6/7, 7/8, 2/21 \\
2003 & October   & 3/4, 15/16, 18/19, 19/20 \\
2003 & December  & 7/8 \\
2004 & February  & 19/20 \\
2004 & April     & 15/16, 19/20, 20/21, 22/23 \\
2004 & May       & 10/11, 12/13, 14/15, 23/24, 24/25 \\
\hline
\end{tabular}}
\end{table}

All images were de-biased, dark current subtracted and flat-fielded
using the IRAF\footnote{IRAF is distributed by the National Optical
Astronomy Observatory, which is operated by the Association of
Universities for Research in Astronomy, Inc., under a cooperative
agreement with the National Science Foundation.} package.
The photometry was extracted with the help of the {\it Difference Image                            
Analysis Package} (DIAPL)\footnote{The package is available at
http://users.camk.edu.pl/pych/DIAPL} written by Wo\'zniak (2000)
and recently modified by W. Pych.
The package is an implementation of the method developed by
Alard and Lupton (1998). A reference frame was constructed
by combining 7 individual images taken during dark time on the
night of May 24/25, 2003. Profile photometry for the reference frame
was extracted with DAOPHOT/ALLSTAR (Stetson 1987).
These measurements were used to transform the light curves
from differential flux units into instrumental magnitudes, which
later were transformed to the standard $R$-band magnitudes by adding
a median offset of 0.549 mag, derived from data on
3894 stars presented by Hatzidimitriou \etal (2004).
Finally, we performed period search and analysis with the TATRY code
(see methods in Schwarzenberg-Czerny 1989, 1996).

\Section{Detected variables}

Searches for variable stars in M56 led to detection of seven objects.
Besides five known variables, two objects, V13 and V14, are new identifications.
The equatorial coordinates of all objects, as well as their angular
distances from the cluster center ($\alpha_{2000}=19^h16^m35\zdot\ups5$,
$\delta_{2000}=+30^{\arcd}11^{\arcm}05^{\arcs}$, Harris 1996)
are listed in Table~2. Only two of the variables, namely V6 and V13,
are located inside the cluster half-mass radius $r_h=1\zdot\arcm16$
(Harris 1996), but neiter of them are located inside the core
radius $r_c=0\zdot\arcm37$ (Harris 1996).
Table~3 gives photometric data on the variables. The $B$-band
magnitudes for four of the variables, namely V1, V3, V5 and V14,
were taken from Hatzidimitriou \etal (2004). Unfortunately, there
is no photometric information on the three remaining objects:
V4, V6 and V13.

In Fig.~1 we present phased as well as time-domain light curves of
the seven detected variables. Fig.~2 shows $R/B-R$ color-magnitude
diagram of M56 based on the data from Hatzidimitriou \etal (2004).
For all detected variables we adopted the average magnitudes
$<R>$ derived from our data. The magnitudes in the $B$ band were
taken from Hatzidimitriou \etal (2004) if they were available.
Note that for the three objects mentioned above there is no color
information.

\begin{table}
\centering
\caption{\small Positions of detected variables in the field of M56}
\medskip
{\small
\medskip
\begin{tabular}{lccc}
\hline
Name & RA(2000.0) & Dec(2000.0) & Distance from \\
     &            &             & the center \\
\hline
V1  & 19\uph16\upm39\zdot\ups33 & -30\arcd12\arcm16\zdot\arcs7 & 1\zdot\arcm53 \\
V3  & 19\uph16\upm37\zdot\ups82 & -30\arcd12\arcm34\zdot\arcs1 & 1\zdot\arcm59 \\
V4  & 19\uph16\upm27\zdot\ups46 & -30\arcd08\arcm20\zdot\arcs7 & 3\zdot\arcm40 \\
V5  & 19\uph16\upm36\zdot\ups58 & -30\arcd08\arcm46\zdot\arcs6 & 2\zdot\arcm32 \\
V6  & 19\uph16\upm35\zdot\ups79 & -30\arcd11\arcm39\zdot\arcs9 & 0\zdot\arcm59 \\
V13 & 19\uph16\upm38\zdot\ups72 & -30\arcd10\arcm59\zdot\arcs1 & 0\zdot\arcm81 \\
V14 & 19\uph16\upm29\zdot\ups83 & -30\arcd12\arcm27\zdot\arcs2 & 1\zdot\arcm97 \\
\hline
\end{tabular}}
\end{table}

\begin{table}
\centering
\caption{\small Photometric data on detected variables in M56}
\medskip
{\small
\begin{tabular}{lccccccc}
\hline
Name &  $B$  & $<R>$ & $\Delta R$ & $B-<R>$ &      $P$      &   Maximum   & Type \\
     &       &       &            &         &               & in $R$-band & \\
     & [mag] & [mag] &   [mag]    &  [mag]  &      [d]      & JD-2450000  & \\
\hline
V1   & 15.50 & 15.10 &    1.07    &  0.40   &   1.510116(8) &   2784.46   & BL Her \\
V3   & 14.65 & 12.08 &    0.78    &  2.57   &  72 ?         &      -      & SR/Irr \\
V4   &   -   & 15.75 &    0.45    &    -    &   0.423723(4) &   2462.50   & RR Lyr \\
V5   & 14.73 & 12.13 &    0.57    &  2.60   & 145 ?         &      -      & SR/Irr \\
V6   &   -   & 12.18 &    1.22    &    -    &  89.70(19)    &   2498.34   & RV Tau \\
V13  &   -   & 13.61 &    0.13    &    -    &  38.96(3)     &   2881.28   & puls? \\
V14  & 16.54 & 15.69 &    0.27    &  0.85   &   0.380795(5) &   2889.29   & RR Lyr \\
\hline
\end{tabular}}
\end{table}

\begin{figure}[htb]
\centerline{\includegraphics[height=120mm,width=120mm]{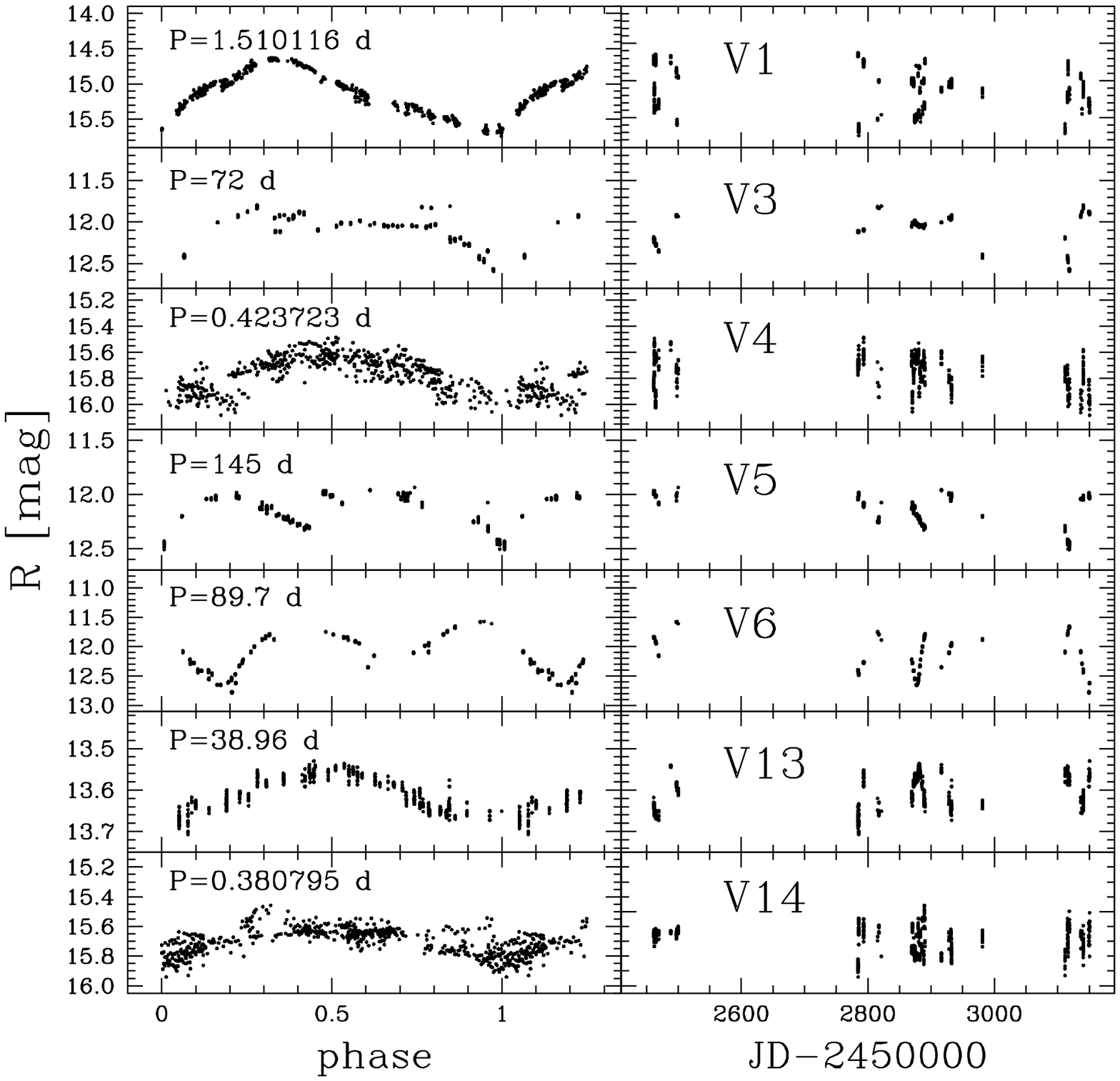}}
\FigCap{
Phased (left column) and time-domain (right column) light curves of
seven variable stars detected in the field of the globular cluster M56.
}
\end{figure}

\begin{figure}[htb]
\centerline{\includegraphics[height=120mm,width=120mm]{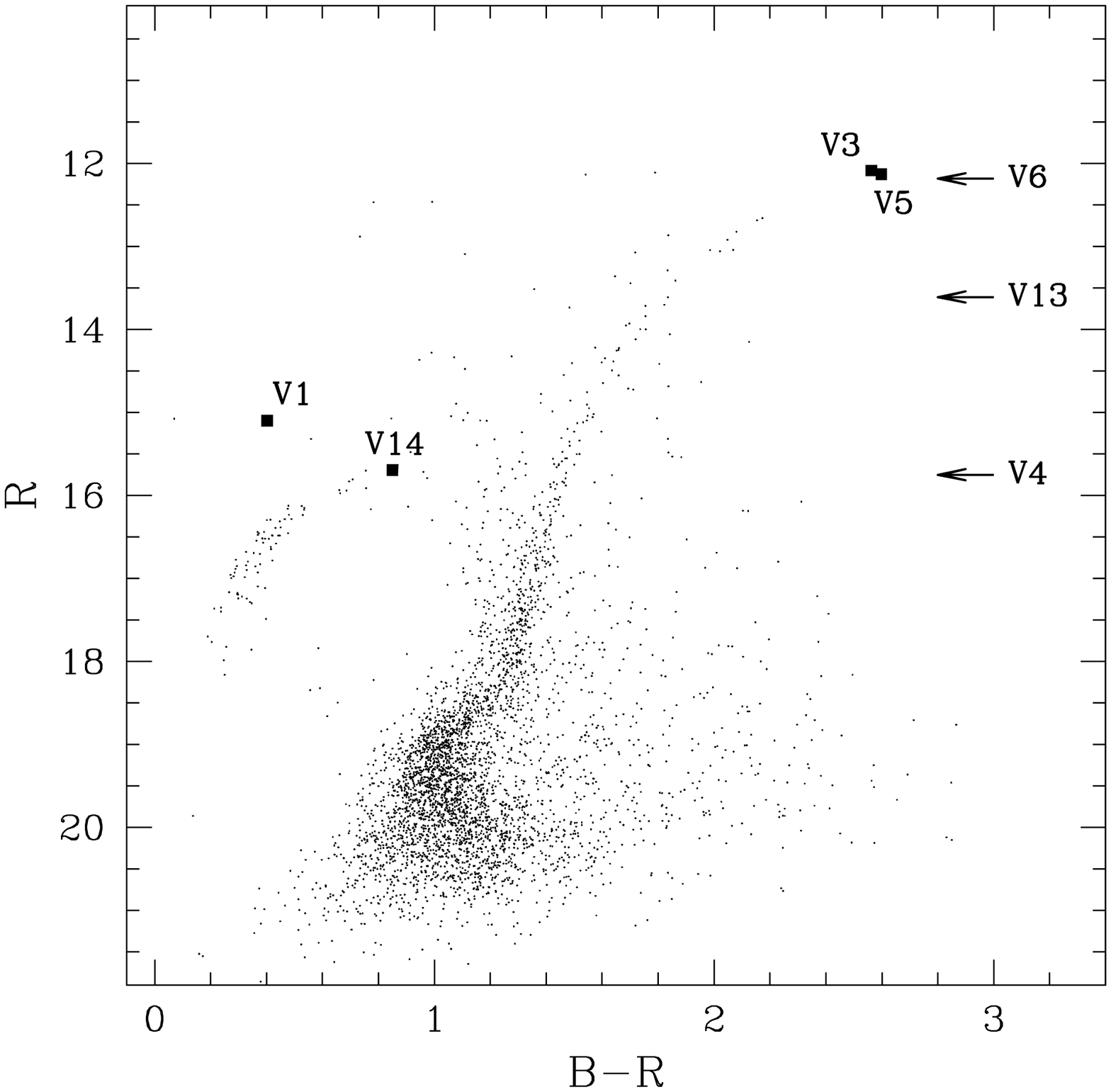}}
\FigCap{
$R/B-R$ color-magnitude diagram of M56. Locations for four
detected variables are marked with squares. For three objects
we only know the average $R$-band magnitudes (those are marked with
arrows).
}
\end{figure}

Variable V1 is a BL Her star whose cluster membership was
proved by a radial velocity study by Harris \etal (1983)
and a relative proper motion study by Rishel \etal (1981).
Based on photographic $B$ and $V$ data obtained in the years
1935--1984 Wehlau and Sawyer Hogg (1985) constructed an $O-C$
diagram. They estimated a secular period change rate in this star
of $(+3.3\pm0.2)$ d/Myr and the period $P_0=1.510019$~d
at the epoch $t_0={\rm JD}2445252.316$.
The different band used in our observations does not
allow us to add another point in that diagram; however we confirm
positive period change in the object. The rate of period change
calculated as
$$
\frac{\Delta P}{\Delta t}=\frac{P_1-P_0}{t_1-t_0}=(+4.4\pm0.4) {\rm ~d}/10^{-6}{\rm yr}
$$
where the values of $P_1$ and $t_1$ come from our data (Table 3), 
is consistent, at three sigma level, with that given by
Wehlau and Sawyer Hogg (1985). Our observations do not show
the presence of neither a dip shortly before maximum light
nor a bump on the descending branch in the light curve,
noticed by them. Their remark could be a result of a small number
of data points per period (only 24 in the $V$-band).

Objects V3 and V5 occupy the same region on the color-magnitude
diagram of the cluster, the tip of the red giant branch. They are
semiregular or irregular variables for which relative proper
motions measured by Rishel et al. (1981) indicate their membership.
The periods of the variables given by Russeva (2000),
namely 42.12 or 34.86~days for V3 and 31.33~days for V5,
does not fit our data at all. We have found other periods in our
data: 72 and 145 days for V3 and V5, respectively, but the
objects require more continuous observations to study
their behavior.

Variable V4 is an RR Lyrae variable and a very likely
member of the cluster, though it is located almost $3r_h$
from the cluster center. Its sine-like light curve
with an amplitude of about 0.45~mag and a period of 0.423723~d
suggest it may be RR Lyrae type ``c''.

Another variable object, V6, is an RV Tau star which belongs
to the cluster (Webbink 1981).
There are only six known such variables in Galactic
globular clusters (Zsoldos 1998). RV Tau stars are pulsating
supergiants of the formal period $P_0$ in the range of 30-150 days,
having spectral types F-G in maximum light
and K-M in minimum light. Their light curves are characterized
by the presence of alternating deep and shallow minima
and the amplitude ratio $A_0/A_1$ of about 1, where
$A_0$ and $A_1$ are the amplitudes of the formal period and
its first harmonic, respectively.

The variable V6 fulfills the classification criteria very well.
Fig.~3 illustrates the power spectrum of the object, calculated
with the help of ANOVA statistics (Schwarzenberg-Czerny 1996).
The two highest peaks at $f_0=0.01115$~c/d and $f_1=0.02236$~c/d
correspond to the formal period and its harmonic, respectively.
In Fig.~4 we show phased light curves of the variable taken
from different epochs from the years 1935--2004.
The presence of alternating minima is obvious. The new data
indicate the formal period of V6 to be $89.70\pm0.19$ days,
slightly shorter than the period published in previous
studies (90.0 days, Sawyer 1949, Wehlau and Sawyer Hogg 1985)
and which was believed to be stable.
We reanalyzed archival data (see Table 4) published in
Wehlau and Sawyer Hogg (1985) and we have found the period
to decrease with a rate of $-0.005\pm0.003$~d/yr.

\begin{figure}[htb]
\centerline{\includegraphics[height=60mm,width=120mm]{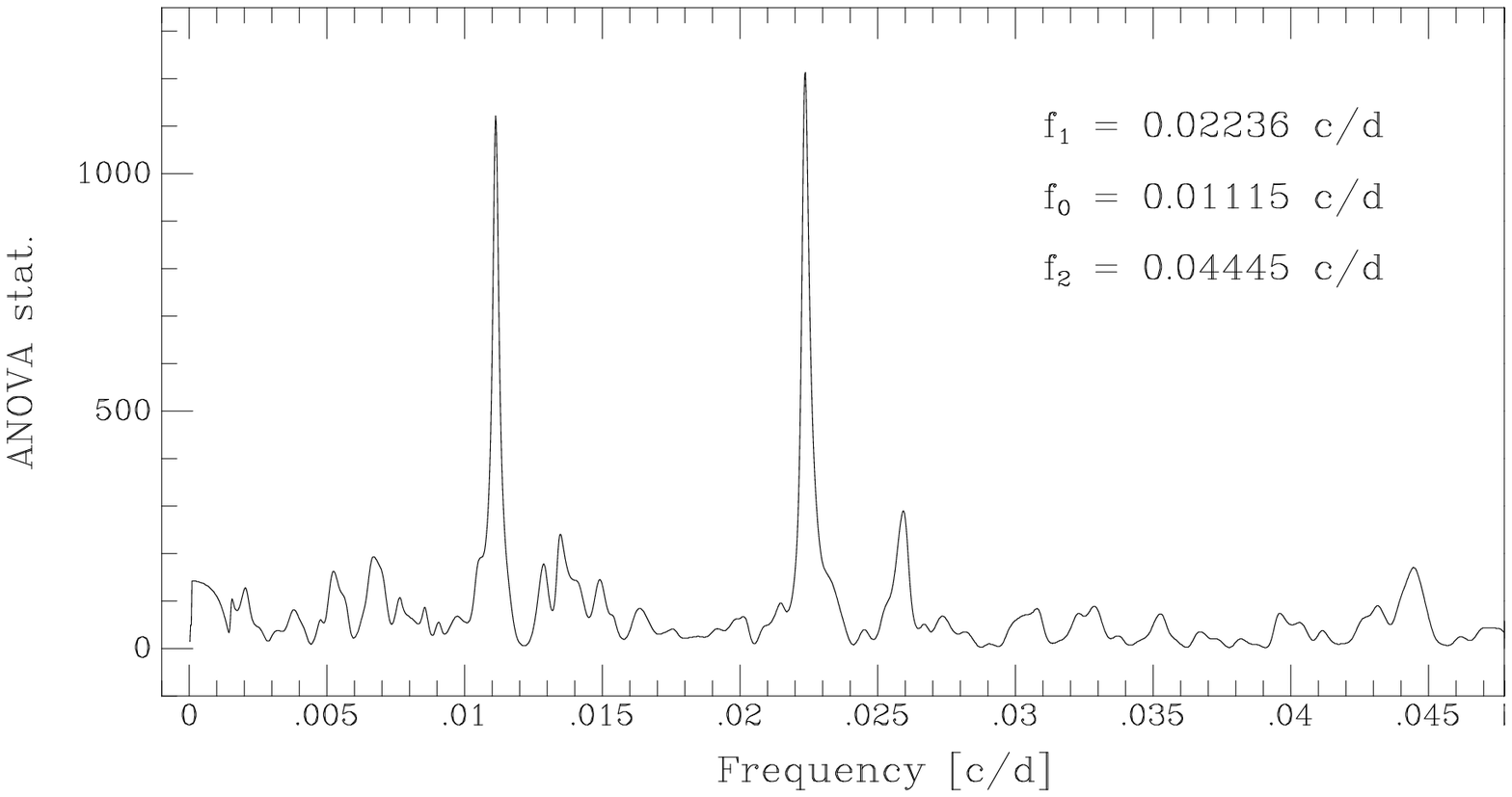}}
\FigCap{The ANOVA power spectrum of the RV Tau star V6.
}
\end{figure}

\begin{table}
\centering
\caption{\small Times of maxima and derived periods of the RV Tau variable V6}
\medskip
{\small
\begin{tabular}{rcccc}
\hline
  E  & Filter & JD$_{max}-2400000.0$ &  $P$  & $\Delta P$ \\
     &        &         [d]          &  [d]  &    [d] \\
\hline
-268 &  $B$   &       28398.75       & 90.06 & 0.02 \\
-114 &  $B$   &       42256.73       & 89.61 & 0.07 \\
   0 &  $R$   &       52498.34       & 89.70 & 0.19 \\
\hline
\end{tabular}}
\end{table}

Variables V13 and V14 are new identifications. The first object
is located inside the half-mass radius of the globular
cluster, but we cannot confirm its membership status.
Unfortunately, there is no color information on this star.
If V13 belongs to M56 it could be an AGB pulsating star.

Variable V14 is the second faintest of the seven detected variables.
It has $R=15.69$~mag on average, an amplitude $\Delta R=0.27$~mag,
and a period of 0.380795~d. These numbers and location of
the star on the color-magnitude diagram are consistent
with classifying V14 as an RR Lyrae-type variable belonging
to the cluster. If this interpretation is true then V14
would be the third RR Lyrae known in the cluster,
after variables V4 and V12.

In this work we confirm, followed by Wehlau and Sawyer Hogg (1985)
and also Russeva (2000), that the object V2 is not variable.
The analysis of our data has not indicated any variability
for the red giant star Kustner 204 suggested by Russeva (2000).
Variables V6-V12 are located outside our field of view, but according
to Wehlau and Sawyer Hogg (1985) only V12, an RRab star of a period
of 0.90608~d, is a member of the globular cluster M56.
Unfortunately, this star is located outside the field of view
of Hatzidimitriou \etal (2004),
making it impossible to place on our color-magnitude diagram.

\begin{figure}[htb]
\centerline{\includegraphics[height=120mm,width=80mm]{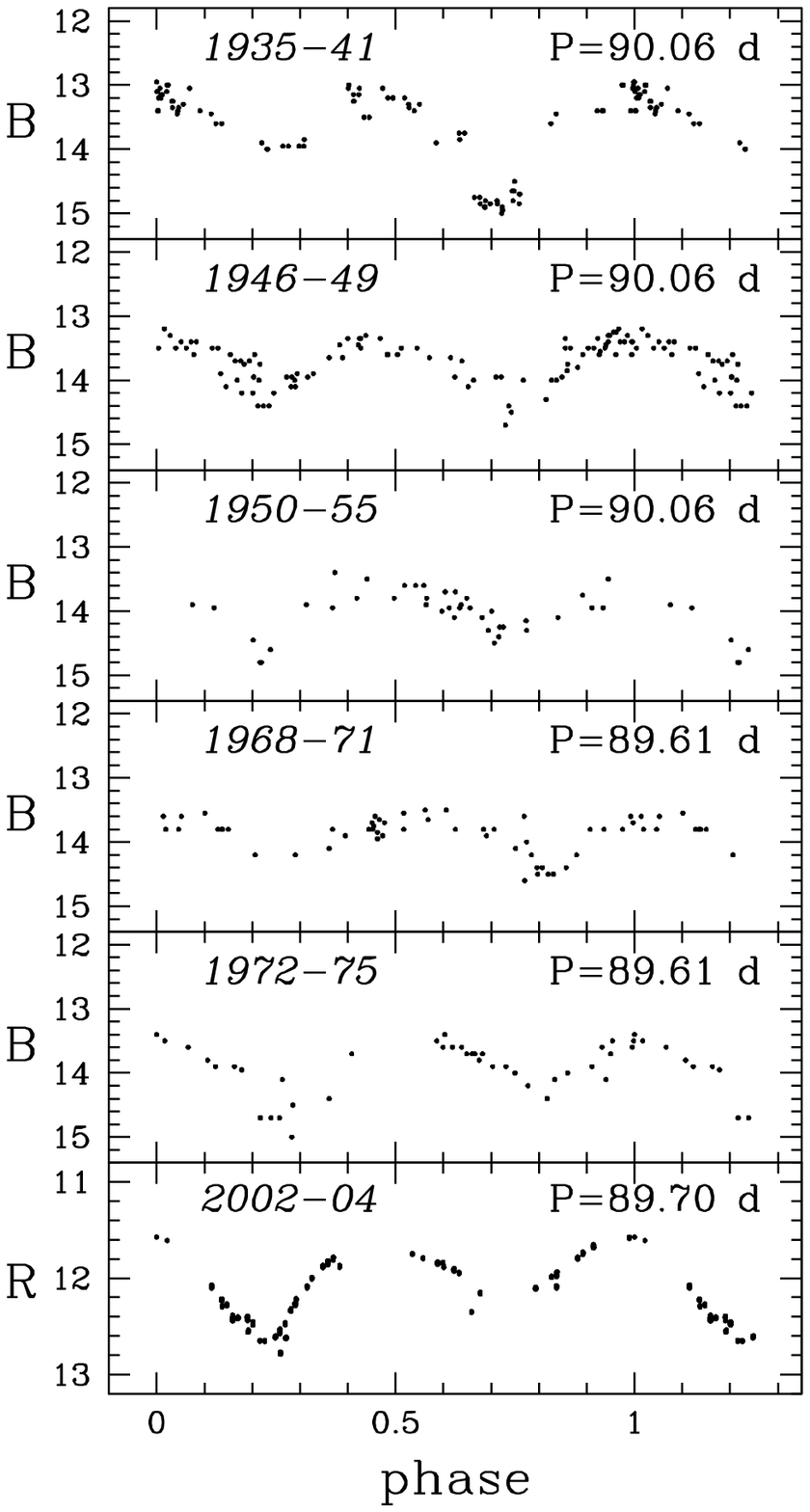}}
\FigCap{Light curves of the RV Tau star V6 in different epochs
from the years 1935--2004. Note the presence of alternating primary
and secondary minima.
}
\end{figure}

\Section{Discussion and Summary}

We have presented the results of a search for variable stars
in the globular cluster M56. Besides five already known variables
we have identified two new objects: V13 and V14.
The object V13 has a period of 38.96 days and probably is
a pulsating AGB star. The variable V14 is very likely an RR Lyrae
star which belongs to the cluster, the third such object in M56,
after V4 and V12. The number of RR Lyrae stars in this
metal-poor globular cluster seems to be very small, but
there are known clusters with similar characteristics.
For example, in M30 of metallicity [Fe/H]$_{CG}=-2.17\pm0.08$
(Carretta 2003) there have been found only 5 such variables
(Clement \etal 2001, Pietrukowicz and Kaluzny 2004); in NGC 6397
of metallicity [Fe/H]$_{CG}=-2.03\pm0.05$ (Gratton \etal 2003)
there is no known RR Lyrae star at all (Kaluzny \etal 2006).

For previously known variables in M56 we have confirmed positive
period changes in BL Her variable V1 and semi-regular nature of
V3 and V5. For variable V6, the RV Tau star, we have found, for the
first time, very likely period change in this star. The negative period change
rate of $-0.005\pm0.003$ d/yr seems to be in contradiction to
the evolutionary status of RV Tau stars as post-AGB objects,
but not with blue-loop evolution. Numerous studies
of period changes in RV Tau (\eg Percy et al. 1997,
Percy and Coffey 2005) also show that $O-C$ diagrams are
dominated by random cycle-to-cycle period fluctuations
of typically 0.005 to 0.02 of a period.
The fluctuations may mask real evolutionary period changes.
Moreover, the interpretation of the diagrams depends on the
specific interval involved.

The results presented here have improved our knowledge
on variable stars in the globular cluster M56, but future searches
will require a bigger telescope (a 1-m or 2-m class telescope)
at an observatory located in a place with better seeing conditions.

\Acknow{
The authors would like to thank Dr. W. Pych for providing some useful
software which was used in the analysis. PP and AO acknowledge support
from the Domestic Grant for Young Scientists of the Foundation for Polish
Science and Polish MNiI grant N203 301 335, respectively. Telescope
operation was supported by the BW grant to Warsaw University Observatory.
}

\end{document}